\documentclass{ws-ijmpe}
\usepackage[super,compress]{cite}
\usepackage{graphicx,subfig,epsfig,epsf,color}
\usepackage{ulem}
\usepackage{amssymb,amsmath,appendix}
\usepackage{slashed}
\usepackage{feynmf}
\usepackage{array}
\usepackage{hyperref}
\begin{document}

\markboth{Debashree Sen, Kinjal Banerjee and T.K. Jha}{Properties of Neutron Stars with hyperon cores in parameterized hydrostatic conditions}

\catchline{}{}{}{}{}

\title{Properties of Neutron Stars with hyperon cores in parameterized hydrostatic conditions}

\author{Debashree Sen\footnote{p2013414@goa.bits-pilani.ac.in}}

\author{Kinjal Banerjee\footnote{kinjalb@goa.bits-pilani.ac.in}}

\author{T.K. Jha\footnote{tkjha@goa.bits-pilani.ac.in}}

\address{Birla Institute of Technology and Science-Pilani, K K Birla Goa Campus, NH-17B, Zuarinagar, Goa-403726, India}

\maketitle

\begin{history}
\received{Day Month Year}
\revised{Day Month Year}
\end{history}

\begin{abstract}

 Models of neutron stars (NSs) with hyperon cores are constructed with an effective chiral model in mean-field approximation. The hyperon couplings are fixed by reproducing their experimentally determined binding energies. The impact of these couplings on population of different particles and the equation of state (EoS) are studied in this work. The global properties of NSs like gravitational mass, radius, baryonic mass and central density are calculated using parameterized Tolman-Oppenheimer-Volkoff equations (PTOV) with special emphasis on two effects of pressure - one contributing to total mass density and the other to self gravity of the star. We find that with PTOV solutions in static conditions, a softer EoS (including hyperons) can also lead to massive stellar configurations of NSs, which are in well agreement with the observed maximum mass bound of $\approx 2 M_{\odot}$ (PSR J0348-0432). Estimates of $R_{1.4}$ and $R_{1.6}$, obtained with the PTOV equations are consistent with the recent findings of the same from the data analysis of gravitational waves (GW170817) observation.
 
\end{abstract}


\ccode{PACS numbers: 97.60.Jd, 26.60.+c, 26.60.c, 14.20.Jn, 04.50.Kd}


\section{Introduction}
\label{intro}

Neutron star (NS) core, composed of degenerate matter, is characterized by very high density (few times the normal matter density $\rho_0 \sim 0.16~\rm{fm^{-3}}$) and extremely low temperature (T = 0 MeV). Under such conditions, the presence of different exotic matter like the hyperons, quarks etc. presents an interesting possibility 
\cite{Glen,Glen2,Glen3,Glen4,Glen5,Glen6}. In this work we study the effects of formation of hyperons ($\Lambda$, $\Sigma^{-,0,+}$ and $\Xi^{-,0}$) on NS properties like mass, radius etc. in an effective chiral model \cite{TKJ,TKJ2,TKJ3,Sen,Sen2}. It is well established that the inclusion of hyperons in neutron star matter (NSM) leads to considerable softening of the equation of state (EoS) and therefore reducing the maximum mass of NSs. However, the discovery of massive NSs like PSR J1614-2230 (M = (1.928 $\pm~ 0.017) M_{\odot}$) \cite{Fonseca} and PSR J0348-0432 (M = (2.01 $\pm~ 0.04) M_{\odot}$) \cite{Antoniadis} gives rise to the ``hyperon puzzle''. Several works, done with phenomenological (using both relativistic or non-relativistic treatments) \cite{Glen,Glen2,Glen3,Glen4,Glen5,Glen6,TKJ,TKJ2,Miyatsu,Stone,Dhiman,Dexheimer,Bednarek, Bednarek2,Weissenborn12,Weissenborn14,Agrawal,Lopes, Oertel,Maslov,Arumugam,Banik,Colucci,Dalen,Lim,Rabhi} and microscopic \cite{Schulze,Baldo,Vidana,Vidana2,Katayama,Yamamoto} approaches, have investigated different ways to solve the ``puzzle''. The hyperons start appearing at densities when the neutron chemical potential exceeds the bare mass of the individual hyperons. But the critical densities of different hyperons depend on the choice of their respective coupling constants. These couplings can be calculated using the well-known schemes like the SU(6) quark model theories \cite{Bednarek,Bednarek2,Weissenborn12,Katayama,Schaffner-Bielich,Sulaksono,Bhowmick,Sahoo} giving the vector and isovector couplings. The scalar couplings are obtained reproducing the binding energies \cite{Glen,Glen2,Glen3,Glen4,Glen5,Glen6,TKJ,TKJ2,Arumugam} of different hyperons in saturated nuclear matter, constrained by certain hypernuclear studies \cite{Glen,Glen5,Rufa}. Some works \cite{Miyatsu,Weissenborn14,Lopes,Weissenborn13,Miyatsu2} have also considered more generalized hyperon couplings  using the SU(3) symmetry theories to study the effect on formation of hyperons in NSM and the global NS properties. 

 NSs are often treated as massive gravity candidates as certain theories of massive gravity, which consider the graviton mass to be non-zero, has been quite successful in describing the properties of NSs, consistent to the observational and empirical bounds on them \cite{Hendi,Sullivan}. It is therefore for such objects like NSs, ordinary General Relativity (GR) may not be a suitable approach. Over a decade several extended/modified gravity theories came up to explain massive gravity \cite{Rham,Brito} and also as alternatives to the dark matter and dark energy theories to explain the total energy budget of the universe. Few such theories also suggest that massive NSs may not only constrain the EoS. They may rather also constrain gravity \cite{Doneva,Eksi,Velten,Oliveira,Brax}. There are several extended/modified theories of gravity like f(R) gravity \cite{Brax,Capozziello,Sotiriou,Arapoglu,Yazadjiev,Staykov,YazadjievPRD91,Santos,Astashenok}, scalar-tensor theories \cite{Brans,Damour,Minamitsuji,Freire,Ramazanoglu,Yazadjiev2,Cisterna,Maselli,Silva,Harada,Novak,Salgado, Sotani,Pani}, quadratic gravity like Einstein-dilaton-Gauss-Bonnet gravity \cite{Yagi,Kleihaus} and Chern-Simons gravity \cite{Alexander,Ali-Haimoud,Yunes,Yagi2}, extended theories of gravity \cite{Wojnar,Wojnar2,Burikham}, massive gravity \cite{Hendi,Sullivan} which are used to modify the general Tolman-Oppenheimer-Volkoff (TOV) equations \cite{tov1,tov2} to calculate the mass and radius of NSs. In certain works \cite{Velten,Glampedakis,Glampedakis2,Schwab} parameterization of the TOV equations has been done for the same purpose, without involving any particular theory of gravity. Refs.\cite{Doneva,Berti:2015itd} (and the refs. therein) recently present commendable reviews of all such theories in this regard. 
  
 In the present work, we evaluate the hyperon couplings, reproducing the binding energies of each the hyperon species ($\Lambda$(-28 MeV), $\Sigma^{-,0,+}$(+30 MeV) and $\Xi^{-,0}$(-18 MeV) \cite{Schaffner-Bielich,Sulaksono,be}). We study the effect of formation of hyperons on the properties of NSs. As discussed earlier, normal GR may not be a suitable approach to calculate the properties of NSs which are highly compact ($M/R \sim 0.2$) and gravitating objects. Therefore we use the parametrized TOV (PTOV) equations as given by \cite{Velten} with two most important parameters in connection to pressure - the one ($\beta$) coupling with inertial pressure to contribute to the overall mass density and the other ($\chi$) resulting in gravitational effects of pressure, known as the self-gravity of the star \cite{Eksi,Schwab}. These parameters and their variance are not arbitrary and each has its own physical significance. We have also constrained the values these two parameters within the framework of our model. This modification/parameterization to normal GR, relevant to the compact objects like NSs, can be applied universally to all EoS to obtain the properties of NS. For example few relativistic mean-field (RMF) models, although they satisfy all the saturated nuclear matter properties, do not fulfill the (2.01 $\pm~0.04) M_{\odot}$ mass constraint of NS mass \cite{Dutra}. With a simple use of these PTOV equations, one may achieve mass as $\sim 2 M_{\odot}$ even for these models. Therefore, it can be said that self-gravity of NSs is equally important as the EoS to contribute to its net gravitational mass. 
  
 The present manuscript is organized in the following way. In section 2 we describe our effective chiral model including the baryon octet. We also specify our model parameters and the coupling scheme adopted for the work. The basic formalism to obtain static properties of NS using both general and parameterized TOV equations are also discussed in this section. The results obtained are shown and discussed in section 3. We also present a possible conservative bound on $\beta$ and $\chi$ necessary to be consistent with the most massive NSs observed. The final conclusions of the work are mentioned in section 4. The results with negative values of $\beta$ and $\chi$ are presented and discussed in Appendix \ref{app_negative}.

\section{Formalism}
\subsection{The Effective Chiral Model with baryon octet}
The effective chiral model is based on the zero temperature relativistic mean field theory. The effective Lagrangian density is given by \cite{TKJ,TKJ2,TKJ3,Sen,Sen2}

\begin{eqnarray}
\mathcal{L} = \overline{\psi}_B \left[\left(i \gamma_{\mu} \partial^{\mu} 
- g_{\omega_B}~ \gamma_{\mu} \omega^{\mu}  -\frac{1}{2} g_{\rho_B}~ 
\overrightarrow{\rho_{\mu}} \cdot \overrightarrow{\tau} \gamma^{\mu} \right)-g_{\sigma_B} 
\left(\sigma + i \gamma_5 \overrightarrow{\tau} \cdot \overrightarrow{\pi} \right) \right] \psi_B \nonumber \\ + \frac{1}{2} \left(\partial_{\mu} \overrightarrow{\pi} 
\cdot \partial^{\mu} \overrightarrow{\pi} + \partial_{\mu} \sigma ~ \partial^{\mu} \sigma \right)
-{\frac{\lambda}{4}} \left(x^2-x_0^2\right)^2 - \frac{\lambda B}{6} (x^2-x_0^2)^3 
- \frac{\lambda C}{8}(x^2-x_0^2)^4 \nonumber \\
-\frac{1}{4}F_{\mu\nu}F^{\mu\nu} +\frac{1}{2}\sum_B g_{\omega_B}^2~x^2~\omega_\mu 
\omega^\mu -\frac{1}{4}~\overrightarrow{R}_{\mu\nu} \cdot \overrightarrow{R}^{\mu\nu}
+\frac{1}{2}~m_\rho^2 ~\overrightarrow{\rho_\mu} \cdot \overrightarrow{\rho^\mu} 
\end{eqnarray}

where, $\psi_B$ is the baryon spinor while $\tau$ and $\gamma^{\mu}$ are the Pauli and Dirac matrices, respectively. The kinetic terms for the $\omega$ and $\rho$ fields are $-\frac{1}{4}F_{\mu\nu}F^{\mu\nu}$ and $- \frac{1}{4}\overrightarrow{R}_{\mu\nu}\overrightarrow{R}^{\mu\nu}$, respectively, where $F_{\mu\nu}=\partial_\mu \omega_\nu - \partial_\nu \omega_\mu$ and $\vec{R}_{\mu\nu}=\partial_\mu \rho_\nu - \partial_\nu \rho_\mu$. The subscript B denotes sum over all baryonic states viz. the nucleons and the hyperons (sumover index $B = n, p, \Lambda, \Sigma^{-,0,+}, \Xi^{-,0}$). $x_0$ is the vacuum expectation value of scalar field due to spontaneous breaking of the chiral symmetry and $x^2 = ({\pi}^2+\sigma^2)$ makes both the scalar and vector fields chiral invariant. 

 The nucleons (N=n,p) and the hyperons (H=$\Lambda$, $\Sigma^{-,0,+}$, $\Xi^{-,0}$) interact with eachother via the scalar $\sigma$ meson, the vector $\omega$ meson (783 MeV) and the isovector $\rho$ meson (770 MeV). The corresponding coupling strengths are $g_{\sigma_B}, g_{\omega_B}, g_{\rho_B}$, respectively. They are evaluated at nuclear saturation density $\rho_0 = 0.153~fm^{-3}$ along with the higher order scalar couplings $B$ and $C$. The explicit contributions of pions are neglected in this case since in the mean field treatment, $< \pi >= 0$ and their mass becomes $m_{\pi}=0$. Hence we neglect their explicit contributions and consider only the non-pion condensed state of matter as in \cite{TKJ,TKJ2,TKJ3}. The baryon mass ($m_B$) and scalar and vector meson masses ($m_{\sigma}$ and $m_\omega$) are given as 

\begin{eqnarray}
m_B = g_{\sigma_B} x_0,~~ m_{\sigma} = \sqrt{2\lambda} x_0,~~
m_{\omega} = g_{\omega_N} x_0~.
\end{eqnarray}

 To account for the asymmetric nuclear matter, we incorporate the isospin triplet $\rho$ mesons. Although it is possible to consider the effect of interaction of the $\rho$ mesons with the scalar and the pseudoscalar mesons similar to the $\omega$ meson and to dynamically generate the mass of $\rho$ mesons similar to that of the scalar and vector mesons, we choose to consider an explicit mass term for the isovector $\rho$ meson similar to what was considered in \cite{Sahu,Sahu1,Sahu2,TKJ,TKJ2,TKJ3,Sen,Sen2}.

 The equation of motion (at T=0) for the fields and the corresponding energy density and pressure of the many baryon system are calculated in relativistic mean field approach \cite{Glen,VolKo,MulSer} as a function of baryon density. 

 At higher density, the chemical potential of the neutrons in $\beta$-stable NSM equals and even surpasses the rest masses of the individual hyperons and we take those hyperons at equal footing with the nucleons. The muons also start appearing when the electron chemical potential exceeds the rest mass of muons. The charge neutrality and chemical equilibrium conditions are essentially to be imposed to ensure a state of minimum energy of a charge neutral NSM. The charge neutrality conditions is as follows

\begin{eqnarray} 
\sum_{B} Q_B~\rho_B + \sum_{l} Q_l~\rho_l = 0 
\end{eqnarray}

where, the suffix B is summed over all nucleons and the hyperons while suffix $l$ denotes sumover all leptonic states. $Q_B$ and $Q_l$ are the charge of baryons and the leptons, respectively. The total baryon density in terms of Fermi momenta $k_B$ is 

\begin{eqnarray} 
\rho = \sum_B \rho_B =\frac{\gamma}{2\pi^2} \sum_{B} \int^{k_B}_0 dk ~k^2  
\end{eqnarray}

The value of the spin degeneracy factor $\gamma$ is 2 for this case and $\rho_l$ is the density of each lepton $l=e,\mu$.

The chemical equilibrium conditions are given as

\begin{eqnarray} 
\mu_B=\mu_n-Q_B \mu_e \label{chem_eq}\\ 
\mu_\mu=\mu_e
\end{eqnarray}

where, $\mu_n$ and $\mu_e$ are chemical potentials of neutron and electron, respectively.

The baryon chemical potential is given by

\begin{eqnarray} 
\mu_B=\sqrt{{k_B}^2+{m^*_B}^2} ~+~g_{\omega_B} ~\omega_0 ~ +~g_{\rho_B} I_{3B}~\rho_{03}
\end{eqnarray} 

where, $I_{3B}$ is the third components of isospin and $\omega_0$ and $\rho_{03}$ are the mean field approximate or vacuum expectation values of $\omega$ and $\rho$ fields, respectively given as

\begin{eqnarray}  
\omega_0 = \frac {\sum\limits_{B} g_{\omega_B} \rho_B}{\Bigl(\sum\limits_{B} g_{\omega_B}^2 \Bigr) x^2}
\label{vector_field}
\end{eqnarray} 

and

\begin{eqnarray} 
\rho_{03}=\sum_{B}\frac{g_{\rho_B}}{m_\rho^2}I_{3B}~\rho_B 
\label{isovector_field}
\end{eqnarray} 

The scalar equation of motion in terms of $Y=x/x_0 = m_B^{\star}/m_B$ is given by

\begin{eqnarray} 
\hspace*{-1.5cm}\sum_B \Biggl[(1-Y^2)-\frac{B}{C_{\omega_N}}(1-Y^2)^2+\frac{C}{C_{\omega_N}^2}(1-Y^2)^3 +2\frac{C_{\sigma_B}~C_{\omega_N}}{m_B^2 ~Y^4} \frac{\Bigl(\sum\limits_{B} g_{\omega_B} \rho_B\Bigr)^2}{\sum\limits_{B} {g_{\omega_B}}^2} - 2 \sum_B \frac{~C_{\sigma_B}~\rho_{SB}}{m_B~ Y} \Biggr]=0
\nonumber \\ 
\label{scalar_field}
\end{eqnarray}

where, the scalar density $\rho_{SB}$ of each baryon is given as
 
\begin{eqnarray} 
\rho_{SB}=\frac{\gamma}{2 \pi^2} \int^{k_B}_0 dk ~k^2 \frac{m_B^*}{\sqrt{k^2 + {m^{*}_{B}}^2}}
\end{eqnarray}

 On the basis of this theory, the equations of state (EoS) i.e. energy density ($\varepsilon$) and pressure ($P$) are evaluated as follows :

\begin{eqnarray} 
\hspace*{-1.5cm}\varepsilon = \frac{m_B^2}{8~C_{\sigma_B}}(1-Y^2)^2-\frac{m_B^2 B}{12~C_{\omega_N}C_{\sigma_B}}(1-Y^2)^3
+\frac{C m_B^2}{16 ~C_{\omega_N}^2~ C_{\sigma_B}}(1-Y^2)^4 +\frac{1}{2Y^2}C_{\omega_N} \frac {\Bigl(\sum\limits_{B} g_{\omega_B} \rho_B\Bigr)^2}{\sum\limits_{B} {g_{\omega_B}}^2} \nonumber \\
\hspace*{-5.5cm} + \frac{1}{2}~m_\rho^2 ~\rho_{03}^2 + \frac{\gamma}{\pi^2} \sum_B \int_{0}^{k_B} k^2 \sqrt{(k^2+{m_B^*}^2)} ~dk 
+ \frac{\gamma}{2\pi^2} \sum_{\lambda= e,\mu^-} \int_{0}^{k_\lambda} k^2 \sqrt{(k^2+{m_\lambda}^2)}~ dk \nonumber \\
\protect\label{EoS1}
\end{eqnarray}


\begin{eqnarray}         
\hspace*{-1.5cm}P = -\frac{m_B^2}{8~C_{\sigma_B}}(1-Y^2)^2+\frac{m_B^2 B}{12~C_{\omega_N}~C_{\sigma_B}}(1-Y^2)^3
-\frac{C~ m_B^2}{16~ C_{\omega_N}^2 C_{\sigma_B}}(1-Y^2)^4
+\frac{1}{2Y^2}~C_{\omega_N} \frac {\Bigl(\sum\limits_{B} g_{\omega_B} \rho_B\Bigr)^2}{\sum\limits_{B} {g_{\omega_B}}^2} \nonumber \\ 
\hspace*{-1cm} + \frac{1}{2}~m_\rho^2 ~\rho_{03}^2 + \frac{\gamma}{3\pi^2} \sum_B \int_{0}^{k_B} \frac{k^4}{ \sqrt{(k^2+{m_B^*}^2)}}~ dk 
+ \frac{\gamma}{6\pi^2} \sum_{\lambda= e,\mu^-} \int_{0}^{k_\lambda} \frac{k^4}{ \sqrt{(k^2+{m_\lambda}^2)}}~ dk \nonumber \\
\protect\label{EoS2} 
\end{eqnarray} 

Here $C_{i_B}=(g_{i_B}/m_i)^2$ are the scaled couplings with $i = \sigma, \omega, \rho$ and $C_{\omega_N}=\frac{1}{x_0^2}$.

 The coupling strength for the $\rho$ meson is obtained by fixing the symmetry energy coefficient $J = 32$ MeV at $\rho_0$ and is given by,

\begin{equation}
J = \frac{C_{\rho_N} k_{FN}^3}{12\pi^2} + \frac{k_{FN}^2}{6\sqrt{(k_{FN}^2 + m^{\star 2})}}
\end{equation}

where $k_{FN}=(6\pi^2 \rho_N/{\gamma})^{1/3}$ is the nucleon Fermi momentum. 

\subsection{The model parameters}

 We choose from ref. \cite{TKJ} the parameters which are well in agreement with almost all the empirical and experimental constraints on nuclear saturation properties like binding energy per nucleon ($B/A$), nuclear incompressibility ($K$), effective mass ($m^*$), symmetry energy coefficient ($J$) and slope parameter ($L_0$) at temperature T=0 and saturation density $\rho_0$. They are presented in table \ref{param}.

\begin{table*}[ht!]
\caption{Parameters of the effective chiral model considered for the present work (adopted from \cite{TKJ3}) are displayed. $C_{\sigma_N}$, $C_{\omega_N}$ and $C_{\rho_N}$ are the corresponding scalar, vector and iso-vector couplings. $B$ and $C$ are the higher order couplings of the scalar field. Other derived quantities such as the scalar meson mass $m_{\sigma}$, the pion decay constant $f_{\pi}$ are also displayed along with the nuclear saturation properties derived for these models.}
{\scriptsize{
\setlength{\tabcolsep}{1.5pt}
\renewcommand{\arraystretch}{1.1}
\begin{center}
\begin{tabular}{cccccccccccccc}
\hline
\hline
\multicolumn{1}{c}{Model}&
\multicolumn{1}{c}{$C_{\sigma_N}$}&
\multicolumn{1}{c}{$C_{\omega_N}$} &
\multicolumn{1}{c}{$C_{\rho_N}$} &
\multicolumn{1}{c}{$B/m_N^2$} &
\multicolumn{1}{c}{$C/m_N^4$} &
\multicolumn{1}{c}{$m_N^{\star}/m_N$} &
\multicolumn{1}{c}{$m_{\sigma}$} &
\multicolumn{1}{c}{$f_{\pi}$} &
\multicolumn{1}{c}{$K$} & 
\multicolumn{1}{c}{$B/A$} &
\multicolumn{1}{c}{$J(L_0)$} &
\multicolumn{1}{c}{$\rho_0$} \\
\multicolumn{1}{c}{ } &
\multicolumn{1}{c}{($fm^2$)} &
\multicolumn{1}{c}{($fm^2$)} &
\multicolumn{1}{c}{($fm^2$)} &
\multicolumn{1}{c}{($fm^2$)} &
\multicolumn{1}{c}{($fm^4$)}&
\multicolumn{1}{c}{} &
\multicolumn{1}{c}{($MeV$)} &
\multicolumn{1}{c}{($MeV$)} &
\multicolumn{1}{c}{($MeV$)} &
\multicolumn{1}{c}{($MeV$)} &
\multicolumn{1}{c}{($MeV$)} &
\multicolumn{1}{c}{($fm^{-3}$)} \\
\hline
NM-I    &6.772  &1.995  &5.285 &-4.274   &0.292    &0.85  &509.644   &139.710  &303  &-16.3   &32(89)  &0.153 \\
NM-II   &7.325  &1.642  &5.324 &-6.586   &0.571    &0.87  &444.614   &153.984  &231  &-16.3   &32(87)  &0.153 \\ 
\hline
\hline
\end{tabular}
\end{center}
}}
\protect\label{param}
\end{table*}

 The values of nuclear incompressibility at saturation density $\rho_0=0.153~\rm{fm^{-3}}$ ($K= (231,303)$ MeV) for the two models are consistent to that suggested by \cite{Dutra2,Stone2} and the symmetry energy coefficient ($J = 32$~MeV), calculated at saturation density for the two models are within their experimental limits \cite{Dutra2,j}. However, the obtained value of the slope parameter ($L_0= (89,87)$ MeV) for the present models are little more than the bound imposed by \cite{l} although recently \cite{Dutra2,BaoAnLi} suggests the value of $L_0 = (25 - 115)$ MeV. Moreover, the recent acceptable values of tidal deformability ($\Lambda_{1.4} \leq 800$) and radius ($R_{1.4} < 13.76$ km) for 1.4 $M_{\odot}$ mass star, are obtained from gravitational wave (GW170817) detection from binary neutron star merger \cite{Abbot}. Recently the co-relation of the symmetry energy slope parameter $L_0$ with $\Lambda_{1.4}$ and $R_{1.4}$ has been strongly developed after the detection of GW170817. It is shown in few of such recent works \cite{Fattoyev,ZhenYuZhu} that these values of $\Lambda_{1.4}$ and radius $R_{1.4}$ can be also satisfied by EoS with slope parameter $L_0\sim80$ MeV. The EoS for symmetric and pure neutron matter with the two models are also in good agreement with the heavy-ion collision data \cite{hic} as shown in \cite{TKJ3}. However, the models yield softer EoS compared to other mean-field models \cite{rmf,rmf2,Dutra2} because of high value of nucleon effective mass (0.85,~0.87)$m_N$ obtained with our models. This in turn affects the properties of NSs.

\subsection{Calculation of hyperon coupling constants}
\label{Couplings}

In order to investigate the formation of hyperons in NSM and their impact on global NS properties, one needs to specify the hyperon couplings $x_{i_H}=g_{i_H}/g_{i_N}$ (where, $i=\sigma,\omega,\rho$ and $H=\Lambda, \Sigma, \Xi$). We fix the scalar coupling constants $x_{\sigma_H}={g_{\sigma_H}}/{g_{\sigma_N}}$ and calculate the vector coupling constants $x_{\omega_H}={g_{\omega_H}}/{g_{\omega_N}}$ by reproducing the binding energies of each hyperon species at saturation density \cite{Weissenborn12,Arumugam}. The hyperon potential depths ${(B/A)_{H}}|_{\rho_0}$ are calculated at saturation density in terms of the nucleon scalar potential $U_S(=g_{\sigma_N} \sigma_0)$ \cite{Sahu} and the vector potential $U_V(=g_{\omega_N} \omega_0)$ \cite{Sahu} using the relation \cite{Weissenborn12,Bhowmick}

\begin{equation}
{(B/A)_{H}}|_{\rho_0} = ~x_{\omega_H} ~ {U_{V}}|_{\rho_0} +~ x_{\sigma_H} ~ {U_{S}}|_{\rho_0}
\label{be}
\end{equation}    
     
 The binding energy (potential depth) of a particular hyperon species in saturated nuclear matter is the difference between the Fermi energy of the lowest level ($k=0$) of the hyperon and the mass of the hyperon, which eventually results in eq. \ref{be}. The values of potential depths of each hyperon species at saturation density, extracted from the experimental studies of energy levels of hypernuclei, are ${(B/A)_{H}}|_{\rho_0}$ = -28 MeV for $\Lambda$, +30 MeV for $\Sigma$ and -18 MeV for $\Xi$ \cite{Schaffner-Bielich,Sulaksono,be}. However, studies based on hypernuclear levels restricts the value of $x_{\sigma H} \leq 0.72$ \cite{Glen,Glen5}. Because there is no relation between ${(B/A)_{H}}|_{\rho_0}$ and $x_{\rho_H}$, we have chosen $x_{\rho_H}$ = $x_{\omega_H}$ since both $\rho$ and $\omega$ mesons have quite close values of mass and also both are responsible for producing short range repulsive forces.

 We study the effect of the couplings on relative population of different baryons and leptons. We also obtain EoS for the core. For the crust part, having low density, the well known Baym-Pethick-Sutherland (BPS) EoS \cite{BPS} is employed. 
 
\subsection{General Relativistic and parameterized hydrostatic equilibrium conditions}
\label{TOV-pTOV}

 Considering NS as undeformed sphere of perfect fluid in static conditions, the properties like the gravitational and baryonic masses, radius and central energy density were derived by Tolman \cite{tov1} and Oppenheimer \& Volkoff \cite{tov2} (TOV) on the basis of hydrodynamic equilibrium between gravity and internal pressure of the star. Those equations (eq. \ref{tov1} and \ref{tov2}) are derived purely on the basis of general relativistic theory.
 
\begin{eqnarray}
\frac{dM}{dr}= 4\pi r^2 \varepsilon
\protect\label{tov1}
\end{eqnarray} 

and

\begin{eqnarray}
\frac{dP}{dr}=-\frac{GM(r)\varepsilon}{r^2}\frac{\Big(1+\frac{P}{\varepsilon}\Big)\Big(1+\frac{4\pi^2 r^3 P}{M(r)}\Big)}{\Big(1-\frac{2GM(r)}{r}\Big)}
\protect\label{tov2}
\end{eqnarray} 
where, $G$ is the gravitational constant and c=1.

 The parameterized TOV equations \cite{Velten} are given as

\begin{eqnarray}
\frac{dM}{dr}= 4\pi r^2 (\varepsilon + \tilde{\sigma} P)
\protect\label{modtov1}
\end{eqnarray} 

and

\begin{eqnarray}
\frac{dP}{dr}=-\frac{G(1+ \alpha)M(r)\varepsilon}{r^2}\frac{\Big(1+ \beta \frac{P}{\varepsilon}\Big)\Big(1+ \chi \frac{4\pi^2 r^3 P}{M(r)}\Big)}{\Big(1- \tilde{\gamma} \frac{2GM(r)}{r}\Big)}
\protect\label{modtov2}
\end{eqnarray}    

where, $\alpha$, $\beta$, $\chi$, $\tilde{\gamma}$ and $\tilde{\sigma}$ are five free parameters, independent of eachother. These parameters and their variations are not arbitrary and have proper physical significance : 
 
\begin{itemize}

\item $\alpha$ signifies effective gravitation and is an important parameter for f(R) gravity \cite{Brax,Sakstein,Sakstein2,Velten}, where its value is taken to be 1/3. 
 
 From the TOV equation (eq. \ref{tov2}), it can be seen that the pressure have two different roles to play \cite{Schwab} - i) the total mass density as $\Big(1+\frac{P}{\varepsilon}\Big)$ and ii) the total gravitational mass $\Big(1+\frac{4\pi^2 r^3 P}{M(r)}\Big)$. Therefore two separate parameters are used to take care of the two effects of pressure :
 
\item $\beta$ parameterizes the inertial pressure contributing to the total mass density which arises from the conservation of the energy-momentum tensor while 
 
\item the parameter $\chi$ is introduced for the gravitational mass of NSs, governed by pressure, giving rise to the self-gravity of the star \cite{Schwab}.
 
\item $\tilde{\gamma}$ relates to the curvature contribution, depending on the geometry of the star, which is a unique feature of GR configuration. $\tilde{\gamma}=0$ is a Newtonian case which yields too high values of mass for NSs, breaking the causality limit \cite{Latt-Prak2007}. This effect is also seen in \cite{Velten}.
 
\item $\tilde{\sigma}$ introduces the effect of gravity to mass function. For neo-Newtonian case, its value is taken to be 3 \cite{Oliveira2}, which is certainly not a complete treatment for NSs. Thus we keep $\tilde{\sigma}=0$ as in normal GR case. As $\tilde{\sigma}$ manifests the essence of gravity to the mass function, therefore many f(R) gravity theories come up with an additional equivalent condition of $\tilde{\sigma}\neq0$ or more complicated forms which couples to even higher order values of pressure, along with $\alpha=1/3$ \cite{Velten}. In this connection, several works \cite{Santos,Wojnar,Wojnar2} also modify the energy-momentum tensor for the purpose of restoring the stability condition of the star.
\end{itemize} 
 
 In normal GR case ($\alpha$,$\beta$,$\chi$,$\tilde{\gamma}$,$\tilde{\sigma}$) have configuration (0,1,1,1,0) to get back the general TOV eqs. \ref{tov1}, \ref{tov2} while the configuration ($\alpha$,$\beta$,$\chi$,$\tilde{\gamma}$,$\tilde{\sigma}$)=(0,0,0,0,0) leads to pure Newtonian case.

 In the present work we study only the two separate effects of pressure, varying only $\beta$ and $\chi$ and keeping the other parameters same as normal GR configuration. Note that the configuration ($\beta$,$\chi$)=(0,0) in this case does not reduce to Newtonian case because the setting of $\tilde{\gamma}=1$ still holds the validity of GR condition since $\tilde{\gamma}$ is related to the curvature of the star, which is very pronounced in case of NSs. Variation of $\beta$ and $\chi$ under such circumstance (i.e, with $\tilde{\gamma}=1$) brings further changes to normal GR conditions in terms of pressure. In this work we intend show that with simple possible variations in the way in which pressure can contribute to the total mass density and total gravitational mass, we can achieve high mass NS configurations even with a soft EoS (including hyperons). For this purpose we vary the parameters $\beta$ and $\chi$ within the extent suggested by \cite{Velten,Schwab}. As the concept of inertial mass density arises from the conservation of energy-momentum tensor, it can be said that the inertial mass density varies proportionally with the force needed by the fluid to withstand gravitational collapse. It is well known that this force is greater in massive NSs. Therefore the parameter $\beta$ (as it couples to the inertial mass density) relates to this binding force of the fluid, which strengthens further for massive NSs. However, one can also incorporate this feature for massive NSs by modifying the form of the energy-momentum tensor itself. Also pressure plays an important role to the the total gravitational mass. The concept of this mass appears due to the change in pressure $dP/dr$ which can be looked upon as the equivalent force that accelerates the fluid away from its geodesic \cite{Schwab}. With massive NSs, the effect becomes more prominent and therefore in such cases the parameter $\chi$ relates to the gravitating effects of pressure and the self gravity of the star. The importance of incorporating these variations in case of high mass NSs configurations will be portrayed in the following section.

\section{Result and Discussions}
\subsection{Neutron Star with baryon octet}
 As discussed in section \ref{Couplings}, the vector coupling constants $x_{\omega_H}$ and the scalar coupling constants $x_{\sigma_H}$ are calculated for the two models. We choose $x_{\sigma_H}=0.70$ (within the bound set on $x_{\sigma_H}$ by \cite{Glen,Glen5}) and calculate corresponding values of $x_{\omega H}$ for the two models using eq. \ref{be}. For both the two models (NM-I and NM-II) $\Sigma$ particles have the maximum value of vector coupling strength compared to that of $\Lambda$ and $\Xi$ particles. This is because $\Sigma$s have positive potential depth while it is negative for both $\Lambda$ and the $\Xi$s. It is therefore expected that $\Sigma$s suffer maximum repulsion in NSM. 

\begin{figure}[!ht]
\begin{center}
\includegraphics[scale=0.6]{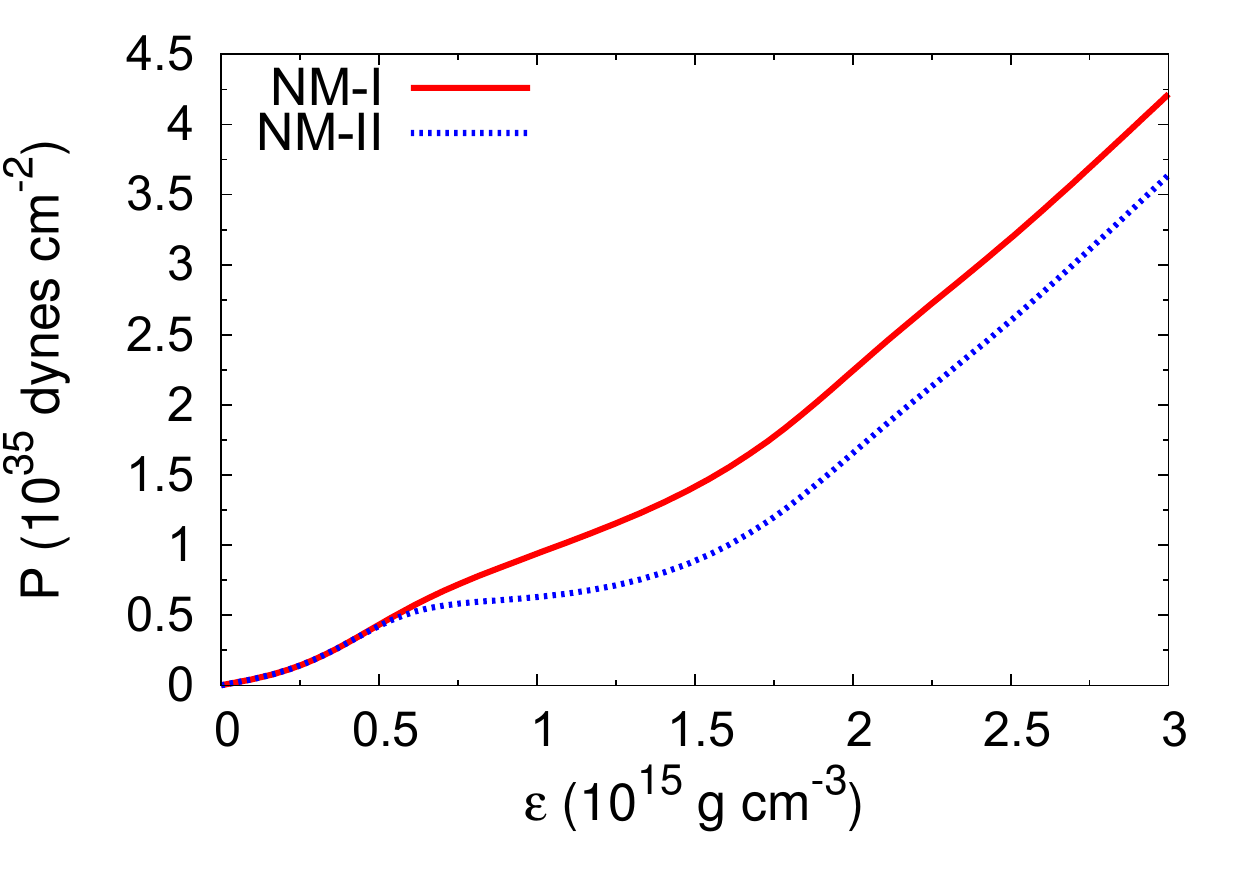}
\caption{\it Variation of pressure with energy density for the two models including hyperons.}
\protect\label{eos_be}
\end{center}
\end{figure}

 The obtained EoS is shown in fig. \ref{eos_be}. NM-I ($K=$~303 MeV) predicts a stiffer EoS than NM-II ($K=$~231 MeV) owing to the larger value of incompressibility $K$ defined at saturation density.
 
\begin{figure}[!ht]
\centering
\subfloat[For NM-I]{\includegraphics[width=0.5\textwidth]{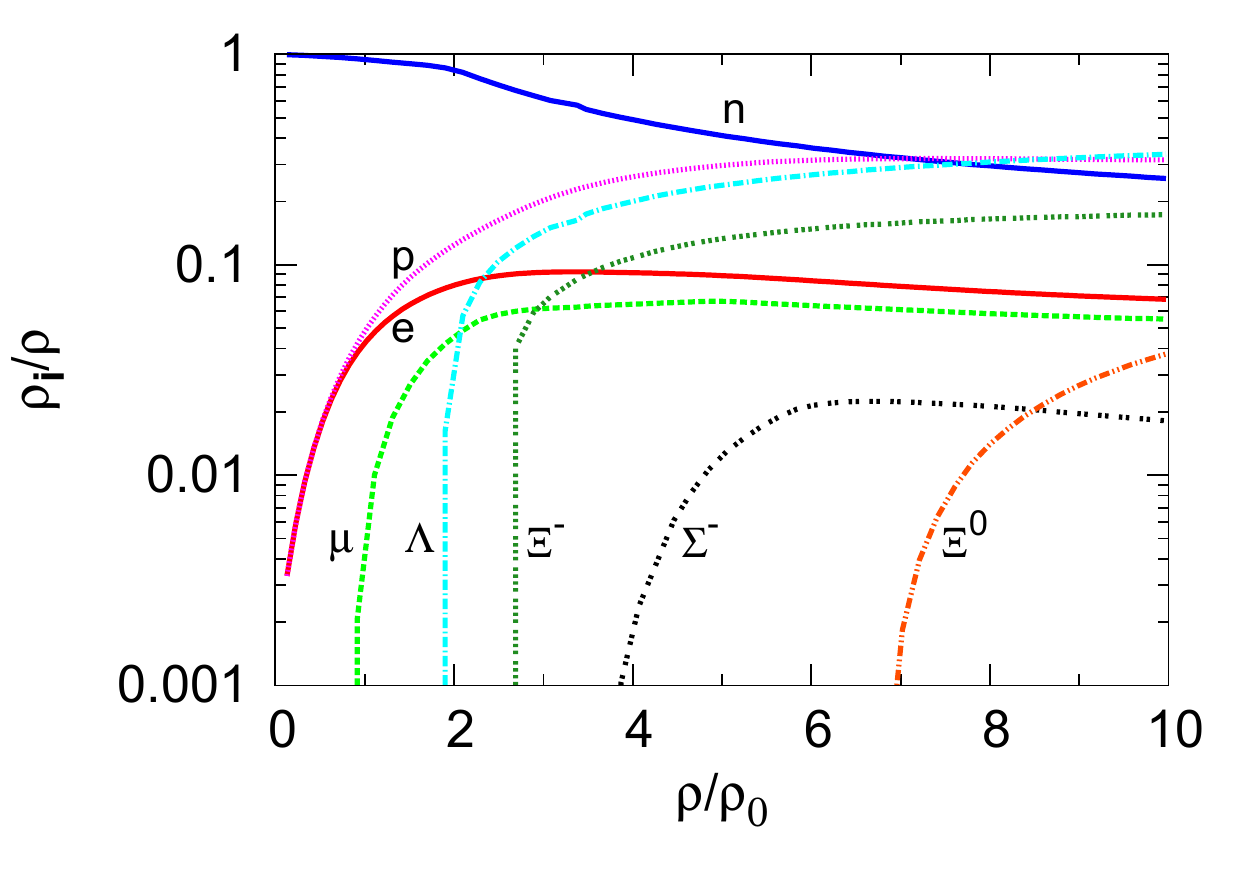}\protect\label{relpop1_be}}
\hfill
\subfloat[For NM-II]{\includegraphics[width=0.5\textwidth]{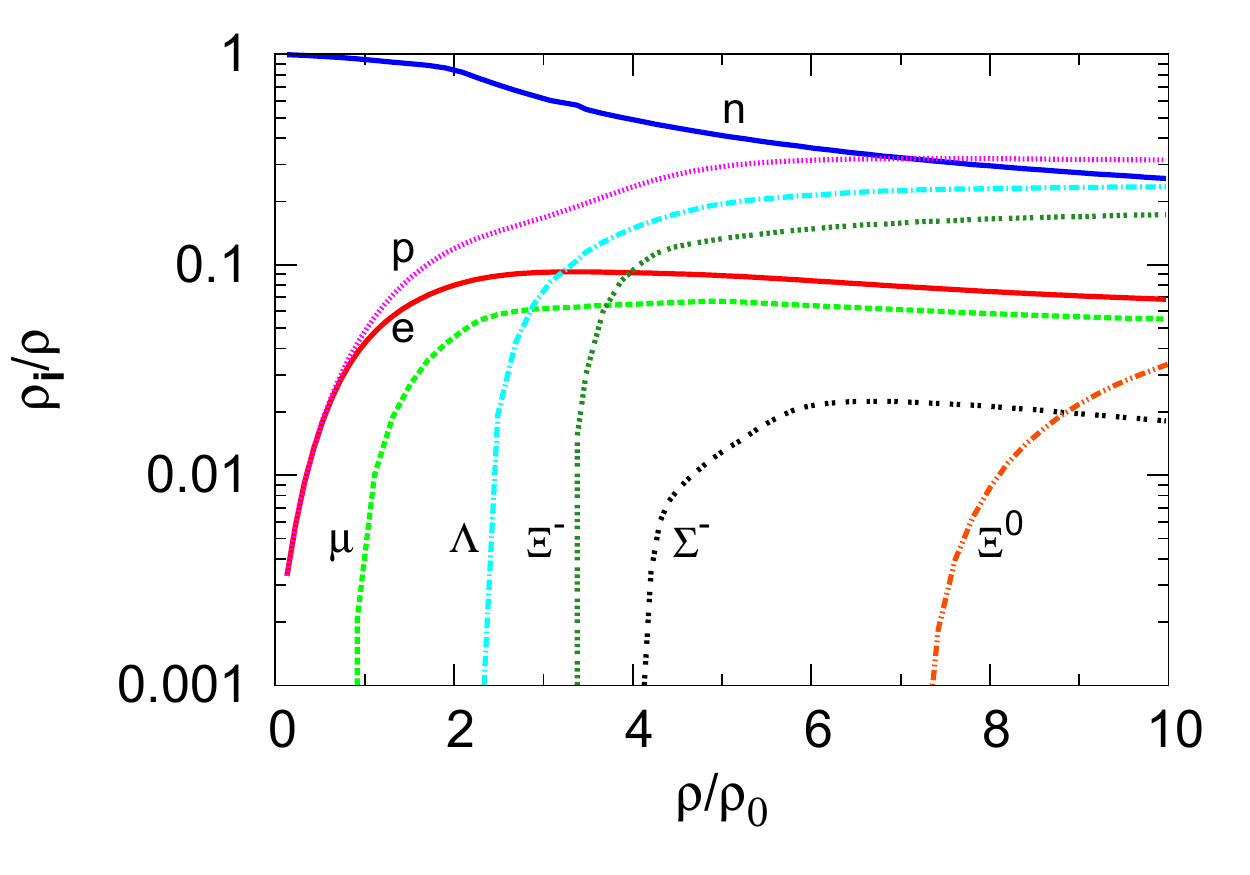}\protect\label{relpop2_be}}
\caption{\it Relative particle population as a function of normalized baryon density.}
\end{figure}

 Figs. \ref{relpop1_be} and \ref{relpop2_be} show the relative population (${\rho_i}/{\rho}$) of different baryons and leptons as a function of normalized baryon density (${\rho}/{\rho_0}$) for models NM-I and NM-II, respectively. Here $i=B,l$. Both the models predict appreciable hyperon population at the expense of neutrons in NSM. For both the models, the lightest hyperon $\Lambda$ appears first (at 1.9$\rho_0$ for NM-I and 2.2$\rho_0$ for NM-II). The next to form is $\Xi^-$ at densities 2.7$\rho_0$ for NM-I and 3.3$\rho_0$ for NM-II, followed by $\Sigma^-$ at 3.9$\rho_0$ (NM-I) and 4.1$\rho_0$ (NM-II). $\Xi^0$ is the last to be formed at 7.0$\rho_0$ for NM-I and at 7.3$\rho_0$ for NM-II.

 For both the models, formation of $\Sigma^0$ and $\Sigma^+$ is totally restricted by their repulsive potential depth. A similar effect has been noted in \cite{Glen6}. However, $\Sigma^-$ appears because the appearance of negatively charged particles are most favored in cold dense matter at densities relevant to NSs as it is easier for a negatively charged baryon to replace a neutron atop Fermi level (according to eq. \ref{chem_eq}) \cite{Glen2}. However, its positive and repulsive potential depth (+30 MeV) makes it appear late even after $\Xi^-$ although it is lighter than the $\Xi$ hyperon \cite{Bednarek,Tolos}. Also it is quite interesting to note from figs. \ref{relpop1_be} and \ref{relpop2_be}, that for both the models $\Sigma^-$ fraction decreases after a certain density (6.7$\rho_0$ for NM-I and 7.2$\rho_0$ for NM-II), unlike the other hyperons. This is because at high densities vector repulsive effects are very high on $\Sigma$s, owing to their positive potential depth in nuclear matter. The deficit in negative charge due to decrease in $\Sigma^-$ is balanced consequently by the high population of $\Xi^-$ and the leptons as seen for both the models. 
 
 Comparing figs. \ref{relpop1_be} and \ref{relpop2_be} we also find that there is feeble shift in critical densities of hyperons (the density at which they start appearing) and difference in concentration with respect to nuclear incompressibility and effective mass \cite{TKJ,TKJ2}. All hyperons appear a bit late and slightly less in concentration in case of NM-II ($K=231$ MeV ; $m_N^{\star}=0.87~m_N$) compared to that in NM-I ($K=303$ MeV ; $m_N^{\star}=0.85m_N$). This difference can be attributed to the difference in effective mass ($m_N^{\star}/m_N$) and nuclear incompressibility ($K$) between the models NM-I and NM-II \cite{TKJ,TKJ2}. Higher value of effective mass (NM-II) the late is the appearance of hyperons. However, as the difference in effective mass between the two models NM-I and NM-II are quite less, there is also feeble shift in onset densities of hyperons. For example the  hyperons start to set in at density 1.9$\rho_0$ for NM-I whereas at 2.2$\rho_0$ in case of NM-II. For the same reason (slightly higher effective mass of NM-II), the hyperon population for NM-II is slightly less than that for NM-I. The effect is most prominent in case of the $\Lambda$s (31\% in case of NM-I and 25\% in case of NM-II) in this work (figs. \ref{relpop1_be} and \ref{relpop2_be}). On the other hand, a lower value of incompressibility (NM-II) yields low baryon chemical potential and softer EoS, therefore, disfavors the early appearance of hyperons. It is therefore the hyperons appear early for NM-I than NM-II. The results are consistent with \cite{TKJ,TKJ2} where the effect of formation of hyperons in light of density and concentration are discussed thoroughly in terms of effective mass ($m_N^{\star}/m_N$) and nuclear incompressibility ($K$).

 The static properties like central energy density ($\varepsilon_c$), gravitational mass ($M$), baryonic mass ($M_B$), radius ($R$) and radius of canonical mass $R_{1.4}$ are calculated subjecting the obtained EoS for the two models to the PTOV equations \ref{modtov1} and \ref{modtov2}. For this purpose we vary only the parameters $\beta$ and $\chi$, as discussed in section \ref{TOV-pTOV} to understand the two separate effects of pressure. The results are tabulated in table \ref{NS_be}. We find that variation in the parameters $\beta$ and $\chi$ bring significant changes to mass of NS, without bringing any huge effect to central density and radius. For different configurations of $\beta$ and $\chi$ in NM-I, the maximum gravitational mass varies from 1.76 ${M_{\odot}}$ to 2.13 ${M_{\odot}}$, maximum baryonic mass changes from 1.96 ${M_{\odot}}$ to 2.25 ${M_{\odot}}$ and corresponding radius 11.2 km to 11.6 km. For NM-II the maximum gravitational mass is found to be (1.65 - 2.05) ${M_{\odot}}$, maximum baryonic mass ranges (1.73 - 2.13) ${M_{\odot}}$ while radius changes (11.0 - 11.4) km. The central energy density is $\approx 1.84\times 10^{15}~ \rm{g~cm^{-3}}$ for NM-I and $\approx 1.48\times 10^{15}~ \rm{g~cm^{-3}}$ for NM-II. The value of $R_{1.4}$ ranges from 12.5 km to 13.2 km for model NM-I with the variations of $\beta$ and $\chi$. For NM-II the same is found to be (11.6 - 12.3) km. $R_{1.6}$ for different values of $\beta$ and $\chi$ ranges as (12.4 - 13.1) km for NM-I and (11.2 - 12.2) km for NM-II.
 
\begin{figure}[!ht]
\centering
\subfloat[For NM-I]{\includegraphics[width=0.5\textwidth]{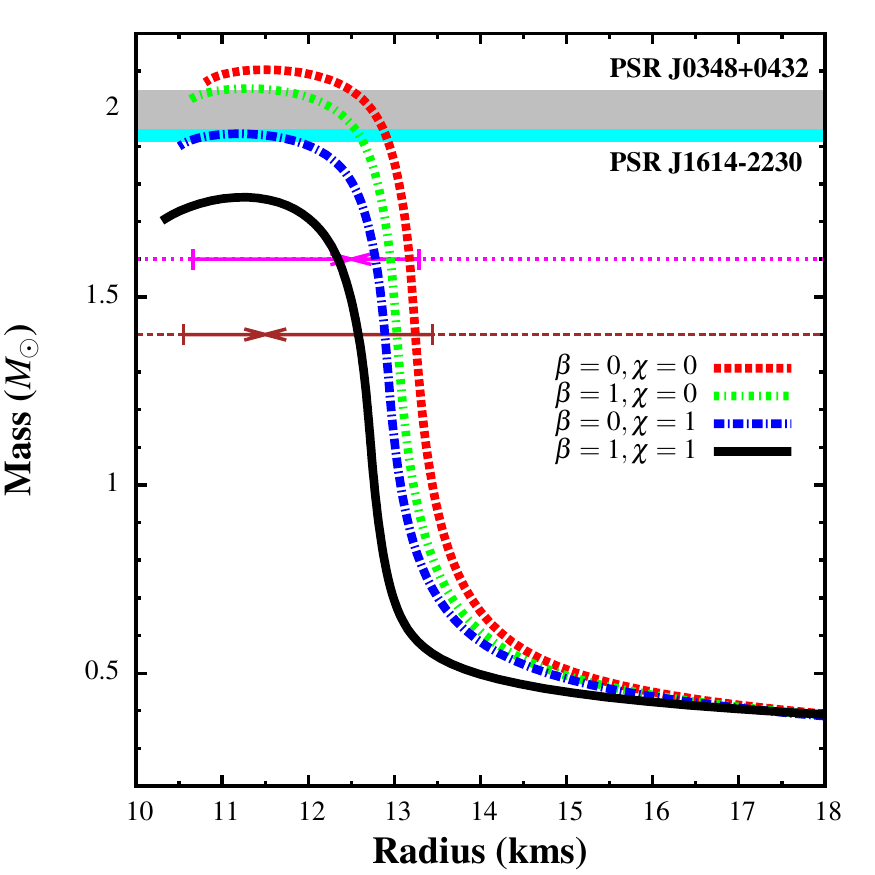}\protect\label{mr_be1}}
\hfill
\subfloat[For NM-II]{\includegraphics[width=0.5\textwidth]{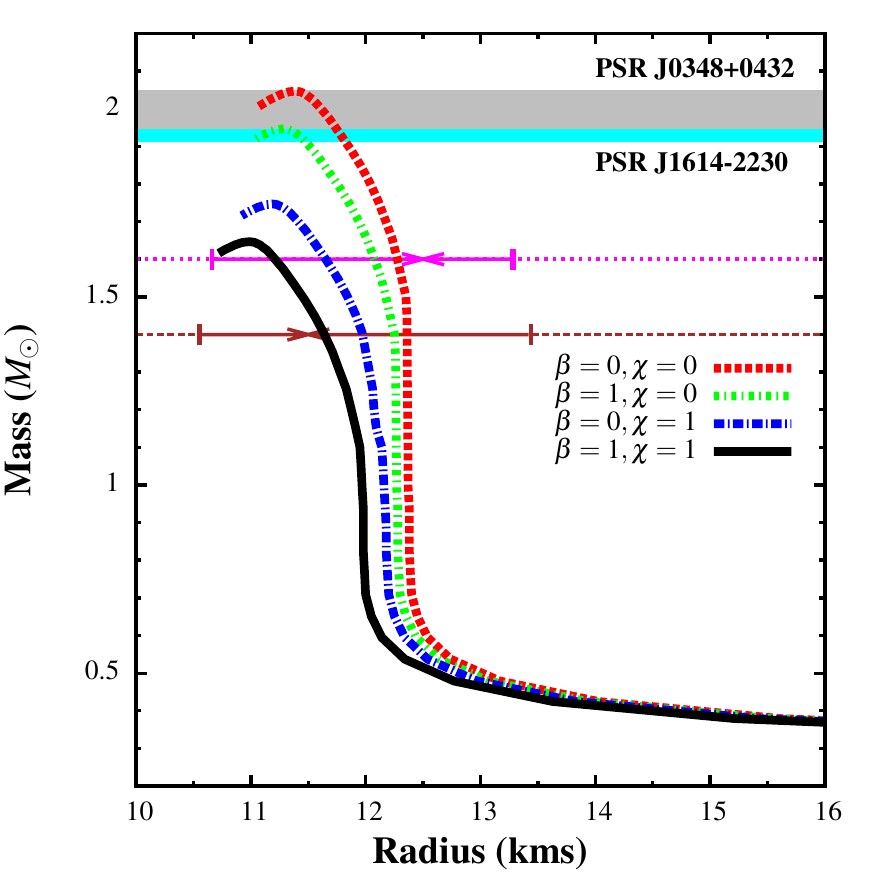}\protect\label{mr_be2}}
\caption{\it Variation of mass with radius for NM-I for different configurations of $\beta$ and $\chi$. The black solid curve shows general TOV solution. Observational limits imposed from high mass stars PSR J1614-2230 ($M = (1.928 \pm~ 0.017) M_{\odot}$) \cite{Fonseca} (cyan band) and PSR J0348-0432 ($M = (2.01 \pm 0.04) M_{\odot}$) \cite{Antoniadis} (grey band) are also indicated. Range of $R_{1.4}$ (brown line) is marked according to \cite{Abbot,Fattoyev} while that of $R_{1.6}$ (magenta line) is marked according to \cite{Abbot,Fattoyev,Bauswein}.}
\end{figure}

\begin{table*}[ht!]
\caption{Static neutron star properties for the models under consideration with respect to the variation in $\beta$ and $\chi$ while $\alpha$,$\tilde{\gamma}$ and $\tilde{\sigma}$ are fixed to normal GR conditions ($\alpha=0,\tilde{\gamma}=1,\tilde{\sigma}=0$). The results from parameterized TOV such as the central density of the star $\varepsilon_c$ (in $\rm{g~cm^{-3}}$), the mass $M$ (in $M_{\odot}$), the baryonic mass $M_B$ (in $M_{\odot}$) and the radius $R$ (in km), radius of canonical mass $R_{1.4}$ (in km) and $R_{1.6}$ (in km) are tabulated.}
\setlength{\tabcolsep}{7pt}
\renewcommand{\arraystretch}{1.1}
\begin{center}
\begin{tabular}{cccccccccccccc}
\hline
\hline
\multicolumn{1}{c}{Model}&
\multicolumn{1}{c}{$\beta$}&
\multicolumn{1}{c}{$\chi$} &
\multicolumn{1}{c}{$\varepsilon_c$} &
\multicolumn{1}{c}{$M$} &
\multicolumn{1}{c}{$M_{B}$} &
\multicolumn{1}{c}{$R$} &
\multicolumn{1}{c}{$R_{1.4}$} & 
\multicolumn{1}{c}{$R_{1.6}$} & \\
\multicolumn{1}{c}{ } &
\multicolumn{1}{c}{ } &
\multicolumn{1}{c}{ } &
\multicolumn{1}{c}{($\times 10^{15} \rm{g~cm^{-3}}$)} &
\multicolumn{1}{c}{($M_{\odot}$)} &
\multicolumn{1}{c}{($M_{\odot}$)}&
\multicolumn{1}{c}{$(km)$} &
\multicolumn{1}{c}{$(km)$} & 
\multicolumn{1}{c}{$(km)$} & \\
\hline
NM-I & 0 & 0 &1.84 &2.13 &2.25 &11.6 &13.2 &13.1\\
     & 1 & 0 &1.83 &2.08 &2.17 &11.3 &13.0 &12.9\\
     & 0 & 1 &1.84 &1.97 &2.13 &11.2 &12.8 &12.7\\
     & 1 & 1 &1.84 &1.76 &1.96 &11.2 &12.5 &12.4\\
\hline
NM-II & 0 & 0 &1.49 &2.05 &2.13 &11.4 &12.3 &12.2\\
      & 1 & 0 &1.49 &1.95 &2.06 &11.3 &12.2 &12.1\\
      & 0 & 1 &1.48 &1.75 &1.87 &11.2 &12.0 &11.6\\
      & 1 & 1 &1.48 &1.65 &1.76 &11.0 &11.6 &11.2\\
\hline
\hline
\protect\label{NS_be}
\end{tabular}
\end{center}
\end{table*}
Figs. \ref{mr_be1} and \ref{mr_be2} shows the variation of gravitational mass with radius for NM-I and NM-II, respectively.

 The maximum gravitational mass, as seen for both the models, is lowest for ordinary GR case (where ($\beta,\chi)=(1,1)$), obtained using the general TOV equations. It is maximum for configuration (0,0) (nearly 22\% more than that obtained with the usual TOV solutions). It is clear that with higher value of inertial pressure contributing to total mass density (i.e. for $\beta=1$), lower is the maximum mass. This is consistent with the results of \cite{Velten}. Same effect is observed for pressure contributing to total gravitational mass (self gravity) i.e. maximum gravitational mass increases with decreasing value of $\chi$. Changes in baryonic mass also follow the same trend as gravitational mass with the change of parameters. For NM-I, configurations (0,0) and (1,0) satisfy the bound on NS mass ($M = (2.01 \pm 0.04) M_{\odot}$) \cite{Antoniadis} from PSR J0348-0432. For NM-II the same constraint is satisfied by configuration (0,0) only. We find that such modifications to normal GR conditions can give rise to maximum mass as high as 2.13 $M_{\odot}$, which is within the cut-off on maximum neutron star mass ($2.0 M_{\odot} < M_{max} < 2.2$ (68 \%) and $2.0 M_{\odot} < M_{max} < 2.6$ (90 \%)) proposed by \cite{Alsing} and \cite{Lawrence}. There is considerable increase in value of canonical radius $R_{1.4}$ with PTOV solutions (maximum 13.1 km for NM-I and 12.2 km for NM-II) compared to that obtained in case of normal GR conditions (12.5 km for NM-I and 11.6 km for NM-II). However, all the values of $R_{1.4}$ are still consistent with that obtained with models like \cite{Stone,Lopes,Dutra,Lat_Lim} and also with the recently suggested value of $R_{1.4}$ by refs. \cite{Abbot,Fattoyev,Most,De}, calculated from the gravitational wave GW170817 observation data of binary neutron star (BNS) merger \cite{Abbot}. The values of $R_{1.6}$, obtained with both the models, are also consistent with the range suggested by \cite{Abbot,Fattoyev,Bauswein} for the same from GW170817.
 
  Overall it can be said that normal GR conditions ($\beta=1$ and $\chi=1$) yield least value of gravitational mass of NSs. To explain massive NS configurations, consistent with observational estimates from PSR J0348-0432 and PSR J1614-2230, values of $\beta$ and $\chi$ must be very low (figs. \ref{mr_be1}, \ref{mr_be2} and table \ref{NS_be}). All the configurations of $\beta$ and $\chi$ used to modify normal GR conditions increase the mass (by 6 - 23 \%) from the normal GR configuration with values of radius and $R_{1.4}$ within the recent bounds prescribed from GW170817 data. It is noteworthy from figs. \ref{mr_be1} and \ref{mr_be2} that the modified effects of pressure on mass density and self gravity of the star are extremely important when one seeks for high mass configurations of NS. At low mass, all the variations of $\beta$ and $\chi$ become gradually ineffective and finally overlap with the normal GR solutions. This effect is also seen in \cite{Velten}.

\begin{figure}[!ht]
\centering
\subfloat[For NM-I]{\includegraphics[width=0.5\textwidth]{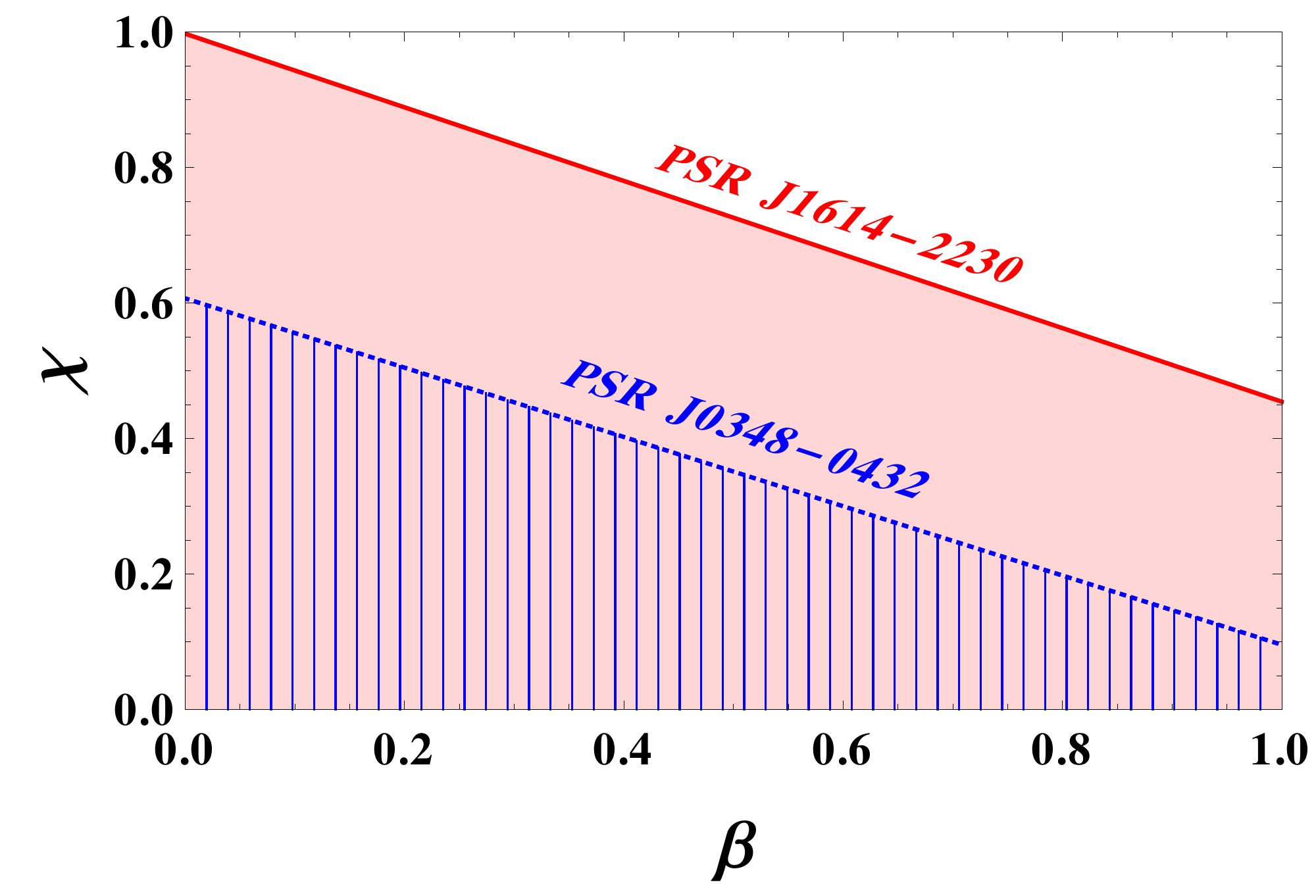}\protect\label{Contour1}}
\hfill
\subfloat[For NM-II]{\includegraphics[width=0.5\textwidth]{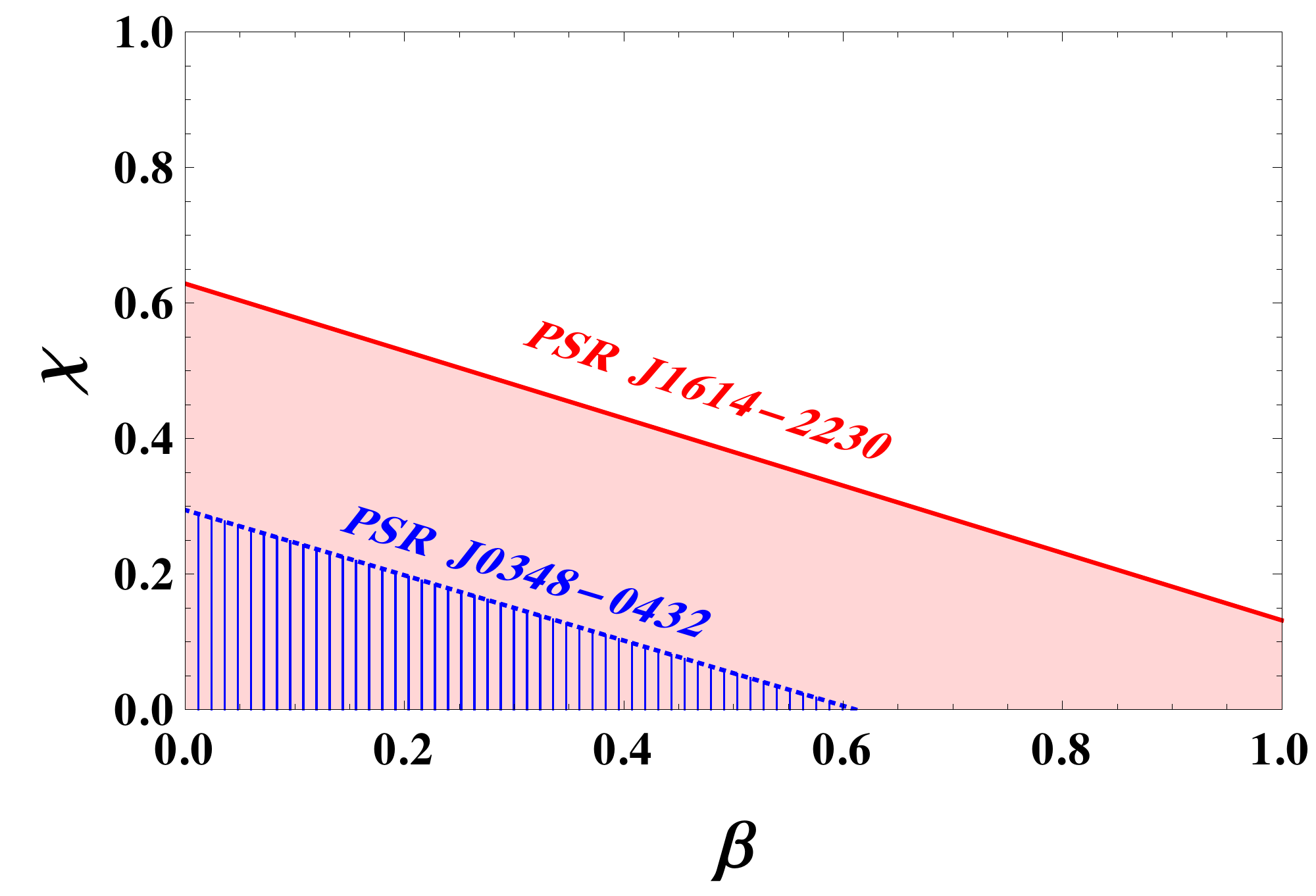}\protect\label{Contour2}}
\caption{\it Contour plot in $\beta$ - $\chi$ plane for gravitational mass constraints from pulsars PSR J1614-2230 ($M = (1.928 \pm~ 0.017) M_{\odot}$) \cite{Fonseca} and PSR J0348-0432 ($M = (2.01 \pm 0.04) M_{\odot}$) \cite{Antoniadis}. The red and blue lines indicate the values of $\beta$ and $\chi$ for which we obtain $M = 1.928 M_{\odot}$ and $M = 2.01 M_{\odot}$, respectively. The red and blue shaded regions represent the allowed values of $\beta$ and $\chi$ to obtain further massive configurations of NSs.}
\end{figure}

 Having obtained the variation of mass with the different configurations of $\beta$ and $\chi$, we now constrain their values with respect to the masses of PSR J1614-2230 and PSR J0348-0432 within the framework of our model. For this purpose we show a contour plot in the $\beta-\chi$ plane in figs. \ref{Contour1} and \ref{Contour2} for NM-I and NM-II, respectively.
 
All the points on the red and blue solid lines in both the figures represent the configurations of ($\beta$, $\chi$) that satisfy the gravitational mass values $M = 1.928 M_{\odot}$ and $M = 2.01 M_{\odot}$, respectively. Therefore these lines indicate the minimum deviation from normal GR case ($\beta=1,\chi=1$) to be taken in order to satisfy the masses of these massive pulsars PSR J1614-2230 and PSR J0348-0432 with our EoS. The red and blue shaded regions indicate all the configurations of ($\beta$, $\chi$) that yield gravitational mass more than $1.928 M_{\odot}$ and $2.01 M_{\odot}$, respectively. It is seen that with a softer EoS (NM-II) (fig. \ref{Contour2}) one requires much more deviation from the normal GR condition than that with a stiffer EoS (NM-I) (fig. \ref{Contour1}) in order to obtain massive NS configurations, consistent with the observational analysis. Also there is quite a considerable difference between the two contour plots shown for NM-I and NM-II in terms of the available space the parameters $\beta$ and $\chi$ can span. With a softer EoS the number of possible combination of ($\beta$,$\chi$) that can yield massive NS configurations are quite less than that in case a stiffer EoS. It is therefore the massive pulsars like 
PSR J1614-2230 and PSR J0348-0432 can be used to constrain the values of these parameters for a given EoS and consequently the allowed space for them in order to yield massive NS configurations do not remain universal but model dependent. Thus we find that although the modified effects of pressure on mass density and self gravity of the star are much essential to obtain high mass configurations of NS, the allowed extent of such modifications still remain dependent on the particular EoS. Existing literature presents a large number of theoretical models suggesting different EoS and composition of NSs which leads to a vast uncertainty to the EoS of NSs. However, given the EoS with our effective chiral model, we obtain well constrained values of the parameters $\beta$ and $\chi$, satisfying the maximum mass criteria of NSs obtained from observational analysis of pulsars PSR J1614-2230 and PSR J0348-0432.

\section{Conclusion}
 In this work we have studied the possible dense matter composition of neutron stars including hyperons. For two model parameter sets, the EoS is obtained with hyperon couplings reproducing the the binding energies of the individual hyperon species. The two models show slightly different results regarding the critical densities and fractions of individual hyperons due to difference in nuclear incompressibility and effective mass. For both models, the population of $\Sigma$ hyperons are very less since they have positive energy (+30 MeV). $\Sigma^-$, though formed, its population density decreases after certain value of density. Models NM-I and NM-II also show difference in gross properties of NS, calculated using both general and parameterized TOV equations. 
 
 One of the important features of the present work is that we show that incorporation of the inertial effects of pressure (via $\beta$) and the self gravity effects (via $\chi$) bring significant changes to the mass of NSs. Also the estimates of $R_{1.4}$ and $R_{1.6}$, obtained using PTOV equations for both model parameters with all configurations of $\beta$ and $\chi$, are consistent with the recent findings of \cite{Abbot,Most,Fattoyev,De,Bauswein} from the data analysis of GW observation (GW170817). Within the framework of our effective chiral model these parameters $\beta$ and $\chi$ are well constrained in terms of the observational masses of massive pulsars PSR J1614-2230 and PSR J0348-0432. Overall it can be said that normal GR conditions ($\beta=1$ and $\chi=1$) yields least value of gravitational mass of NSs. The importance of the modifications made to the normal TOV equations lies in the fact that all the configurations of $\beta$ and $\chi$ used to modify normal GR conditions increases the mass of NS than that obtained with normal GR configuration ($\beta=1$ and $\chi=1$) and help satisfy the observational constraints on high mass of pulsars ($\approx 2~ \rm{M_{\odot}}$) \cite{Antoniadis} and the maximum mass cut-off \cite{Alsing,Lawrence} in some cases. These features are thus very important when one seeks massive NS configurations. The different choice of $\beta$ and $\chi$ show that their values must be low in order to explain massive NS configurations. However, the limiting values of $\beta$ and $\chi$ to satisfy the most massive pulsar masses \cite{Fonseca,Antoniadis} and also the bound on maximum gravitational mass \cite{Alsing} are model dependent. It appears that self gravity has more pronounced effect on the gravitational mass because high mass configurations are obtained only when $\chi=0$. Although the model yields soft EoS on inclusion of hyperons, we are able to satisfy the 2 $\rm{M_{\odot}}$ mass criterion of NSs in static conditions with certain configurations of $\beta$ and $\chi$. It is seen that massive NSs may also constraint gravity instead of only the EoS and self-gravity may play an important role in determining the correct mass of NSs when a particular EoS is considered. However, at present there is a large amount of uncertainty pertaining to the EoS obtained from various models. Therefore the extent to which the massive NSs can constrain gravity still remains inconclusive and a model dependent finding.  
 
\section*{Acknowledgement} 

D.S. thank Atanu Guha for useful discussions and advice regarding certain tools.

\appendix
\section{Static Properties of Neutron Stars with negative values of $\beta$ and $\chi$}
\label{app_negative}

\begin{figure}[!ht]
\centering
\subfloat[For NM-I]{\includegraphics[width=0.5\textwidth]{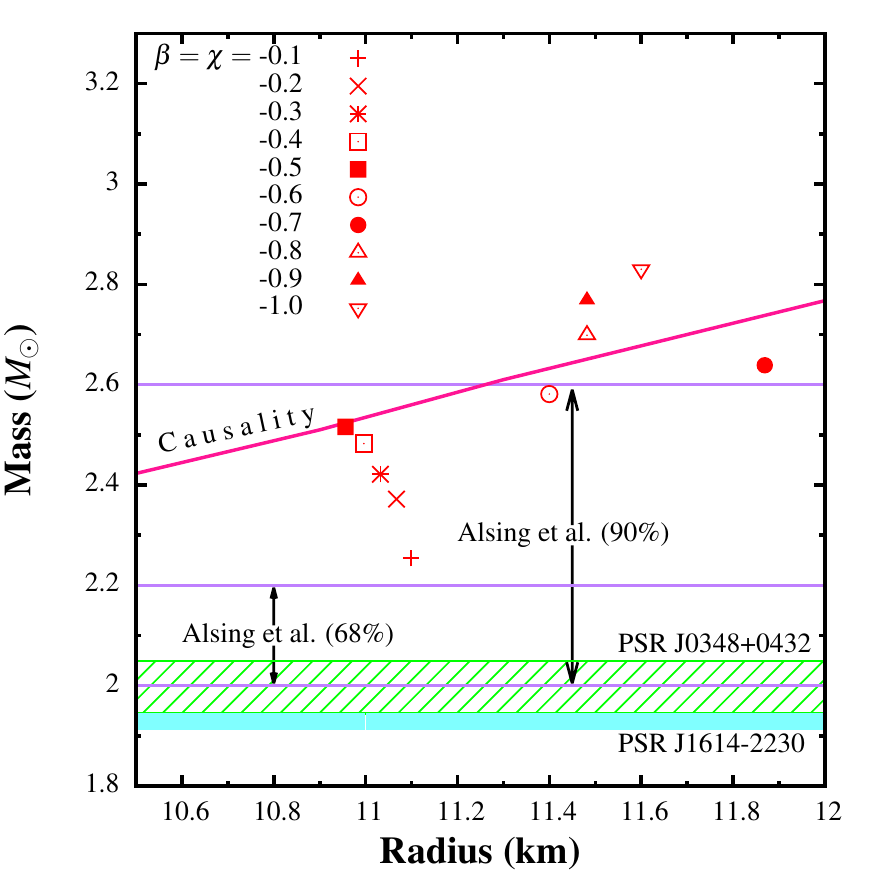}\protect\label{neg1}}
\hfill
\subfloat[For NM-II]{\includegraphics[width=0.5\textwidth]{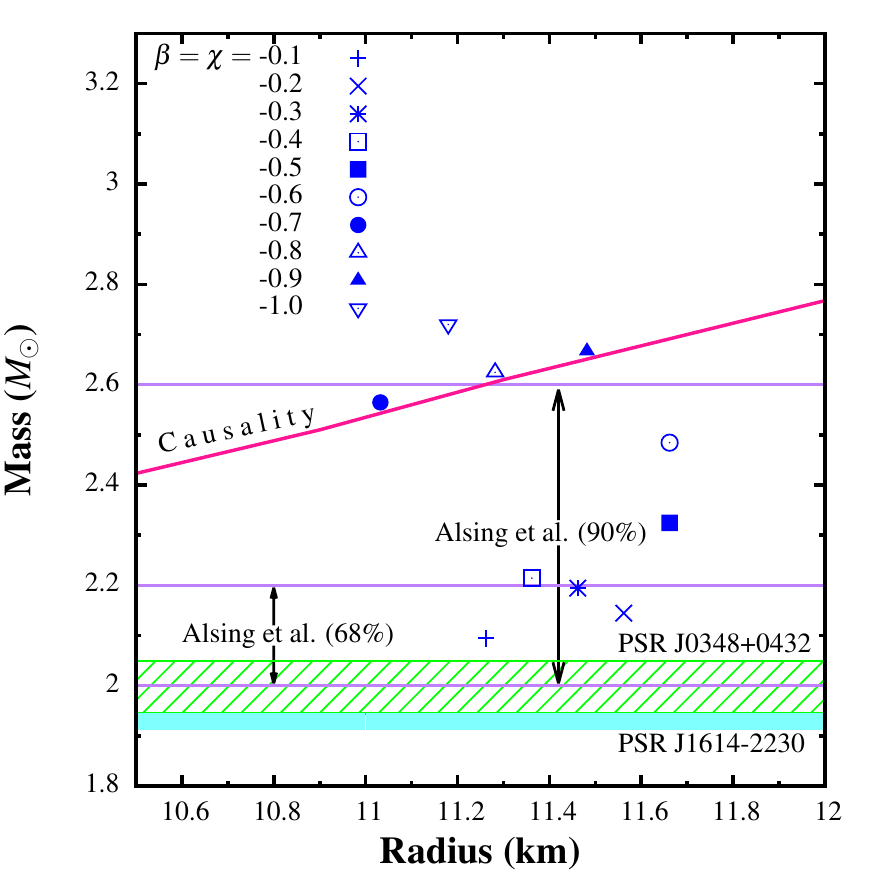}\protect\label{neg2}}
\caption{\it Variation of maximum mass with corresponding radius for the two models with different negative values of $\beta$ and $\chi$. Observational limits imposed from high mass stars PSR J1614-2230 ($M = (1.928 \pm~ 0.017) M_{\odot}$) \cite{Fonseca} (cyan bands) and PSR J0348-0432 ($M = (2.01 \pm 0.04) M_{\odot}$) \cite{Antoniadis} (green bands) are indicated. Constraints on maximum mass from ref. \cite{Alsing} are also shown (areas enclosed by purple lines, indicated by arrows). The causality limit from ref. \cite{Latt-Prak2007} on the $M-R$ plane is also shown (magenta lines).}
\end{figure}

 The inward self-gravitating pressure (parameterized by $\chi$) of the star is balanced by its outward inertial pressure (parameterized by $\beta$) to obtain hydrostatic stability of the stable, non-collapsing, compact NSs. Therefore choosing either of $\beta$ and $\chi$ as negative implies that both these pressures act in the same direction, leading to hydrostatic instability. However, choosing both negative may still render the directions of these two pressures opposite to eachother but the maximum gravitational mass obtained is too high ($M = 2.83 M_{\odot}$ and $R=11.6$ km for NM-I and $M = 2.72 M_{\odot}$ and $R=11.2$ km for NM-II) with such a configuration ($\beta=-1$,~$\chi=-1$). Consequently, for ($\beta=-1$,~ $\chi=-1$) the causality limit on the $M-R$ plane \cite{Latt-Prak2007} is violated and also the maximum gravitational mass estimates are inconsistent with the recent bounds specified for the same in ref. \cite{Alsing} for both the models. However, the value maximum gravitational mass decreases with decreasingly negative values of $\beta$ and $\chi$. We find that for NM-I the configurations $\beta=\chi=$~(-0.1 $-$ -0.6) yield the values of maximum gravitational mass of NSs which are consistent with the range specified by ref. \cite{Alsing}. For NM-II the same is satisfied by the configurations $\beta=\chi=$~(-0.1 $-$ -0.7). Therefore for NM-I the limiting value of $\beta=\chi$ to satisfy the allowed range of maximum mass \cite{Alsing} is -0.6 with corresponding maximum mass $M = 2.58 M_{\odot}$ and radius $R=11.4$ km. Similarly, for NM-II the limiting value of $\beta=\chi$=~-0.7 for the same with corresponding maximum mass $M = 2.56 M_{\odot}$ and radius $R=11.0$ km. In figs. \ref{neg1} and \ref{neg2} we show the values of maximum mass and corresponding radius with different negative values of $\beta$ and $\chi$ for the models NM-I and NM-II, respectively.

\end{document}